\documentclass{article} 
\usepackage{nips13submit_e,times}
\usepackage{comment}
\usepackage{url}
\usepackage{graphicx}
\usepackage{amssymb,amsmath,mathbbol}
\usepackage{rotating}

\newcommand{\ignore}[1]{}

\title{Sparse Predictive Structure \\ of Deconvolved Functional Brain Networks}

\author{
T. Furlanello\thanks{Corresponding authors} ${\;}{}^1$, 
M. Cristoforetti${}^{2,3}$,
C. Furlanello${}^4$, and 
G. Jurman$^{*}{}^4$
\\
\begin{tabular}{c}
${}^1$ CIMEC, University of Trento, Italy\\
${}^2$ ECT$^\star$, Villa Tambosi, I-38123 Villazzano (Trento), Italy\\
${}^3$ LISC, Via Sommarive 18, I-38123 Povo (Trento), Italy\\
${}^4$ Fondazione Bruno Kessler, Trento, Italy
\end{tabular}
\\
\begin{tabular}{c}
\texttt{\{tfurlanello,cristoforetti\}@gmail.com},
\texttt{\{furlan,jurman\}@fbk.eu}\\
\end{tabular}
}

\nipsfinalcopy 

\begin{document}
\maketitle
\begin{abstract}
  The functional and structural representation of the brain as a
  complex network is marked by the fact that the comparison of noisy
  and intrinsically correlated high-dimensional structures between
  experimental conditions or groups shuns typical mass univariate
  methods. Furthermore most network estimation methods cannot
  distinguish between real and spurious correlation arising from the
  convolution due to nodes' interaction, which thus introduces
  additional noise in the data. We propose a machine learning pipeline
  aimed at identifying multivariate differences between brain networks
  associated to different experimental conditions. The pipeline (1)
  leverages the deconvolved individual contribution of each edge and
  (2) maps the task into a sparse classification problem in order to
  construct the associated "sparse deconvolved predictive network",
  i.e. a graph with the same nodes of those compared but whose edge
  weights are defined by their relevance for out of sample predictions
  in classification. We present an application of the proposed method
  by decoding the covert attention direction (left or right) based on
  the single-trial functional connectivity matrix extracted from
  high-frequency magnetoencephalography (MEG) data. Our results
  demonstrate how network deconvolution matched with sparse
  classification methods outperforms typical approaches for MEG
  decoding.
\end{abstract}

\section{Introduction}
\label{sec:intro}
Brain Networks \cite{bullmore2009} (BNs) are high-dimensional objects
that represent the pairwise statistical dependence of measured brain
signals. In this framework, sensors or regions of interests are
represented as nodes and their interactions as links with an
associated weight.
One of the most critical drawback of this approach is that the
construction of a BN results into a fully-connected graph where
spurious and non spurious associations cannot be
distinguished. Furthermore, it is typically of interest to study the
statistical pattern of associations underlying a specific Cognitive
State, but functional patterns found in the data are typically
confoundable with those representing default brain functioning
(RIF). Methods for the estimation of sparse graphs, \emph{i.e.}
networks where only a subset of the links have a non-zero weight, have
been implemented by either using $l1$ penalized regularization
techniques \cite{meinshausen2006high,friedman2008sparse}
or by thresholding the network. However, the amount of sparsity
depends on a regularization (or thresholding) parameter whose choice is completely
arbitrary unless the network estimation is integrated in a
classification framework \cite{zanin2012optimizing}. Recently machine
learning techniques have been extended to the classification of f-MRI
state-dependent Brain Networks
\cite{richiardi2011decoding,shirer2012decoding} as well as for the
classification of patients \cite{bassett2011altered} from resting
state data, revealing an intrinsic relationship between the BN
structure and both transient and permanent cognitive states (see
\cite{Richiardi2013} for an extensive review). Given the high
dimensionality of BN data, PCA-based dimensionality reduction
techniques \cite{leonardi2013principal,schluep2013principal} 
and recursive feature selection methods based on node summary
statistics \cite{fekete2013} or topological features
\cite{jietopological, jie2012structural} have been proposed. In this
paper we adopt the predictive classification framework and consider a
sparsified approach (the $l1l2$ improvement of the elastic network) in
integration with a recent network deconvolution method
\cite{feizi2013network}  for detecting direct effects from an observed correlation matrix containing both direct and indirect effects. The algorithm removes the effect of all indirect paths of arbitrary length in a closed-form solution by exploiting eigen-decomposition. We present the pipeline in
Section~\ref{sec:meth}. In Section~\ref{sec:apps}, we compare such an
off-the-shelf approach with published methods as well as with SVM and
Random Forest predictors on the MEG Biomag 2010 competition 1 dataset
and discuss results in Section 4.  We show that the new procedure can
be used to produce sparse BNs that are predictively associated with
the experimental conditions of interest, with an advantage in terms of
predictive accuracy as well as of interpretability.
\section{Methods}
\label{sec:meth}
\textbf{Notation} Let $E=\{X,Y\}$ be the experiment of interest
composed by $N$ trials, each associated with a label $y_{n}\in
\{1,-1\} $, occurring $N_{1}$ and $N_{-1}$ times respectively, with $N_{1}+N_{-1}$ $n\in N$. Brain
signals are recorded from $p$ loci with a temporal sampling of $t$
data points per trial, resulting in the $p\times t$ data matrix
$X_{n}$.

The goal of our pipeline is that of approximating the frequency specific BNs
$ \phi^{fr}(X)$ and the function $f: \phi^{fr}(X)\mapsto Y$.  Let $ \phi^{fr}(X_{n})$ be
the $p \times p$ matrix whose $(i,j)-th$ element $ \phi^{fr}(X_{n})_{ij}$
corresponds to the spectral coherence $c^{fr}_{ij}$ at frequency $fr$
between the $i-th$ and $j-th$ row of the matrix $X_{n}$. Since the matrix
$ \phi^{fr}(X)$ is symmetric, $\hat{f}$ is defined over a $\frac{p (p-1)}{2}$
vectorial space, where abusing notations $ \phi^{fr}(X_{n})$ is used for both
vectors in such a space and the $p \times p$ matrix representation of
the same vectors. Let $^{d}\phi^{fr}(X_{n})$ indicate the deconvolved network
and $D$ the deconvolution filter. For the case of interest let the
approximated function be linear in the form $\hat{f}= \phi^{fr}(X)\beta+\mu$
such that $\beta\in \textrm{argmin}
\sum_{i=1}^{n}(Y_{i}- \phi^{fr}(X_{n})\beta^{fr})^{2}$, and let $\beta^{fr}_{ij}$ be the
weight of the linear functional $\beta^{fr}$ associated to element
$ \phi^{fr}(X)_{ij}$.

\textbf{Data Analysis Pipeline}
\begin{enumerate}
\item \textbf{The experiment} $E=\{X,Y\}$ is randomly split in two parts $S$
  times keeping the classes balanced: the development
  $E^{dev}_{s}=\{Y^{dev}_{s},X^{dev}_{s}\}$ and test
  $E^{te}_{s}=\{Y^{te}_{s},X^{te}_{s}\}$.
\item \textbf{Network Construction}: Each data input matrix $X_{n}$ is
  transformed into $\phi^{fr}(X_{n})$ by estimating the spectral
  coherence between each combination of rows in $X_{n}$ at frequency
  $fr$.
\item \textbf{Network Deconvolution}: Each network $\phi^{fr}(X_{n})$
  is deconvolved using a non-linear spectral filter $D: \phi^{fr}(X)\mapsto$$
  ^{d}\phi ^{fr}(X)$. Following \cite{feizi2013network} networks are
  deconvolved in three steps: first the network $ \phi^{fr}(X)$ is linearly
  scaled such that all its eigenvalues are between -1 and 1, then the
  rescaled network is decomposed with SVD and finally its eigenvalues
  are transformed such that
  $\lambda^{d}_{i}=\frac{\lambda_{i}}{\lambda_{i}+1}$, with
  $\lambda^{d}_{i}$ is the $i-th$ eigenvalue of the deconvolved
  network and $\lambda_{i}$ the $i-th$ eigenvalue of the rescaled
  network. As shown in \cite{feizi2013network}, under the assumption
  that the observed links are formed by an infinite sum of indirect
  dependencies from nodes at increasing distances, such deconvolution
  procedure leads to the optimal solution.
\item \textbf{Two Steps Elastic Net}: the Elastic Net function
  $f^{fr}_{s}$ is estimated for each split at each frequency. The
  Elastic Net considered here is a linear regression with mixed $l1l2$
  norm, introduced by \cite{zou2005} as a solution for the Lasso
  \cite{tibshirani1996regression} instability problems in $p\gg n$
  environment with highly correlated variables. Since the $l1$ norm
  introduces a shrinkage effect biasing the estimated coefficients, we
  followed \cite{DeMol2009} and considered an elastic net(EN) estimation
  in two phases by first minimizing using the proximal algorithm of \cite{mosci2010solving}\footnote{\url{http://www.disi.unige.it/person/MosciS/CODE/PASPAL.html}} the function
\begin{equation}
\label{l1l2}
 (\sum_{1}^{n} Y_{i}-^{d}\phi^{fr}(X_{i})\beta^{fr}) ^{2}+\tau \|\beta^{fr}\|^{1} +\mu\|\beta^{fr}\|^{2} 
\end{equation}
followed by a debiasing step where a ridge regression is estimated
using only the variables selected (indicated by $\phi^{fr}(X_{i})^{\beta^{fr}\neq 0}$ )in the previous step  by the
minimization of:
\begin{equation}
\label{l2}
 (\sum_{1}^{n} Y_{i}-^{d}\phi^{fr}(X_{i})^{\beta^{fr}\neq 0}\beta^{fr}) ^{2} + \lambda\|\beta^{fr}\|^{2} 
\end{equation}
The parameters $\tau ^{*}$ and $\lambda^{*}$ are tuned using a k-cv
over $E^{d}_{s}$ while $\mu$ is chosen to be small and constant across
tuning. Only the model at the optimal frequency $f^{fr^{*}}$ associated with the highest k-cv accuracy is selected.
\item \textbf{Test Accuracy Estimation } Each $f^{fr^{*}}_{s}$ is tested on their correspondent test set $E^{te}_{s}$.
\item \textbf{Sparse Deconvolved Predictive Network} Results are
  aggregated in the predictive network $B^{fr}$ with entries
  $B^{fr}_{ij}=|\frac{\sum_{1}^{S}\beta^{fr}_{ij}}{S}|$.
\end{enumerate}
\section{Application to MEG Single Trial Classification}
\label{sec:apps}
We apply our method to the MEG Biomag 2010 competition 1 dataset
\cite{van2009attention}, which being among the first publically
available\footnote{\url{ ftp://ftp.fcdonders.nl/pub/courses/biomag2010/competition1-dataset}}
benchmarks for MEG decoding has been extensively considered for
testing novel algorithms and pipelines
\cite{bahramisharif2010covert,signoretto2012classification,kia2013discrete},
sometimes stumbling in selection bias and severely over-estimating
the classification accuracy \cite{olivetti2010}. The experiment
consists in monitoring the brain activity of the subjects in a MEG
scanner under two different conditions: attention had to be covertly
modulated either on the right visual field or on the left
one. Subjects had to fix a cross at the center of the screen and at
regular intervals a cue indicated which direction they had to
covertly
 attend the designated visual field during the next
2500ms. The
 competition data consists in the MEG measurements of
four subjects
 from 500ms before the cue offset to 2500ms after from
274 DC SQUID
 axial gradiometers sensors downsampled at 300Hz. The
goal of the
 competition is to classify on what visual field is the
subject
 modulating his attention based on the MEG data. For our
application we restrict our attention to the first subject and use a
total of 126 trials per condition, for a total of 252 trials to
guarantee balanced stratification between conditions.

\textbf{Preprocessing, Network Construction and Deconvolution} The raw
signals of each trial are independently decomposed with a multitaper
frequency transformation in the 5-40  Hz interval with 2  Hz bin
width. The results of the frequency transforms are used to construct a
coherence network for each trial, which is successively rescaled such
that its eigenvalues are between +1 and -1. After rescaling, the
eigenvalues of the network are filtered as explained in
Section~\ref{sec:meth} \cite{feizi2013network}.

\textbf{Elastic Net Model Estimation} The dataset is randomly divided
10
 times in a class-balanced development (168 trials) and test
(84 trials) splits. The reported classification results are the
average prediction accuracy on the test set of the models estimated
on the correspondent development set. Inside each development split,
the optimal parameters $\tau^{*}$,$\lambda^{*}$ of the Elastic Net
are tuned through grid search selecting those with
 higher prediction
accuracy. Finally the $\beta^{fr}$ weights are computed with the
optimal parameters.

\textbf{Sparse Predictive Network}
The optimal results of each development-test split are aggregated by averaging the $\beta^{fr}$ weights. This procedure leads to the construction of a Sparse Predictive Network which can be either directly analyzed or used to filter the average networks of each class for results interpretation. 

\textbf{Competing Methods}
In order to test the quality of the proposed method we used
other two well established learning algorithms for comparison purposes: Random Forest (RF) \cite{breiman2001random}, based on a growing a set of decision trees, and Support Vector Machine (SVM) \cite{cortes1995support}\footnote{implemented in mlpy\cite{albanese2012mlpy} \url{http://mlpy.sourceforge.net/}}, which is instead looking for a geometric separation surface. Moreover, we tried three different versions of the SVM: the classical linear approach (L-SVM), a kernel based on the Hamming distance \cite{hamming1950error} (H-SVM) and the novel IM graph kernel (IM-SVM), based on the Ipsen-Mikhailov network metric \cite{ipsen2002evolutionary,jurman2011biological,jurman2012}.
As a major difference, both RF and SVM Linear and SVM Hamming work on the vectorized nets, so they are making no use of the network structure, which instead is a key feature when using the IM kernel, which is the Radial Basis Function matrix stemming from the distance matrix computed between all pairs of involved nets \cite{cortes2003positive}. We refer to \cite{kia2013discrete} to compare our results to a pipeline using the Elastic Net directly on summary statistics of spatio-temporal activation patterns and on their Discrete Cosine Transformations.

\section{Results and Conclusions}
\textbf{Non Deconvolved Networks Results}
Given the 4-cv, consistently with previous findings\cite{van2009attention} on this dataset, we found that alpha (8-13 Hz) is the optimal interval. As non-deconvolved Brain Networks at 11 Hz exhibits higher 4-cv accuracy and lower standard errors, we decided therefore to focus our experiments only on those networks.
As shown in Table \ref{table:acc}, the performance of all the classifiers based on local properties of raw Brain Networks are comparable or superior to the 0.67 accuracy of the low level features that \cite{kia2013discrete} employs as baseline. Non-surprisingly RF and EN reach higher results, probably because of their embedded feature selection. Instead, the IM-SVM kernel only reaches near chance-level results, implying that  the topological features of the graph are not useful for classification in this task. We speculate that this may be caused by the symmetric nature of the task.

\textbf{Deconvolved Networks Results}
As an effect of network Deconvolution all the local methods except for RF increase their predictive power 1) strongly bridging the gap between methods with and without embedded feature selection 2) outperforming the 0.67 benchmark and all reaching comparable results with 2-D DCT basis. Furthermore the combination of both the network based Elastic Net and the SVM with Hamming Kernel with Network Deconvolution reaches an accuracy of 0.74 slightly outperforming the 2-D DCT Elastic Net method.

\textbf{Conclusions}
We provide interesting results on the possibility to efficiently use Deconvolved Brain Networks to represent high frequency neural time series from MEG data in decoding problems. Despite the recent increase of application of graph Kernels based on the graph isomorphism paradigm \cite{vishwanathan2010graph,jietopological,jie2012structural,mokhtari2012decoding,vega2013brain}, our comparison of different classification methods suggests that in a partially symmetric system like the brain it is fundamental to consider both local and global characteristics as it may be the case that topological properties do not convey useful informations. Future improvements of our pipeline will be devoted to its adaptation to temporal and multiplex networks whose mathematical formalism has been recently extended in \cite{zhou2010time,holme2012temporal,de2013mathematical,sole2013spectral,kivela2013multilayer}. Finally we notice how the graphical form of the solutions might allow better interpretability and understanding of the brain multivariate structure with the application of common complex network tools, like node level metrics \cite{rubinov2010} or higher level properties like communities \cite{girvan2002community,newman2004finding,porter2009communities} or core-peripheries \cite{borgatti2000models, holme2005core,rombach2012core,bassett2013task} to cross-validated predictive sub-networks.
\begin{table}
\centering
\caption {Average Classification Accuracies for Deconvolved and non Deconvolved 11 Hz Networks, standard errors in brackets. As a baseline \cite{kia2013discrete} reach 0.67 accuracy using the elastic net with summary statistics of spatio-temporal activations and 0.73 using 2-D DCT basis.\\}
  \begin{tabular}{l|c|c|}
  ~       & Non-Deconvolved & Deconvolved\\ \hline 
  L-SVM & ~ 0.65 (0.02)           & 0.72 (0.02) ~          \\
  H-SVM & ~ 0.67 (0.03)           & \textbf{0.74 (0.03)} ~          \\
	IM-SVM & ~  0.56 (0.04)     & 0.48 (0.05) ~          \\  
  RF & ~  0.71 (0.01)     & 0.70 (0.02) ~          \\
  EN  & ~     0.71 (0.03)    & \textbf{0.74 (0.03)} ~          \\
  \end{tabular}
  
  \label{table:acc}
\end{table}
%

\bibliographystyle{unsrt}
\bibliography{furlanello13sparse_arxiv}

\begin{thebibliography}{10}

\bibitem{bullmore2009}
Ed~Bullmore and Olaf Sporns.
\newblock Complex brain networks: graph theoretical analysis of structural and
  functional systems.
\newblock {\em Nature Reviews Neuroscience}, 10(3):186--198, 2009.

\bibitem{meinshausen2006high}
Nicolai Meinshausen and Peter B{\"u}hlmann.
\newblock High-dimensional graphs and variable selection with the lasso.
\newblock {\em The Annals of Statistics}, 34(3):1436--1462, 2006.

\bibitem{friedman2008sparse}
Jerome Friedman, Trevor Hastie, and Robert Tibshirani.
\newblock Sparse inverse covariance estimation with the graphical lasso.
\newblock {\em Biostatistics}, 9(3):432--441, 2008.

\bibitem{zanin2012optimizing}
Massimiliano Zanin, Pedro Sousa, David Papo, Ricardo Bajo, Juan
  Garc{\'\i}a-Prieto, Francisco del Pozo, Ernestina Menasalvas, and Stefano
  Boccaletti.
\newblock Optimizing functional network representation of multivariate time
  series.
\newblock {\em Scientific reports}, 2, 2012.

\bibitem{richiardi2011decoding}
Jonas Richiardi, Hamdi Eryilmaz, Sophie Schwartz, Patrik Vuilleumier, and
  Dimitri Van De~Ville.
\newblock Decoding brain states from fmri connectivity graphs.
\newblock {\em Neuroimage}, 56(2):616--626, 2011.

\bibitem{shirer2012decoding}
WR~Shirer, S~Ryali, E~Rykhlevskaia, V~Menon, and MD~Greicius.
\newblock Decoding subject-driven cognitive states with whole-brain
  connectivity patterns.
\newblock {\em Cerebral cortex}, 22(1):158--165, 2012.

\bibitem{bassett2011altered}
Danielle~S Bassett, Brent~G Nelson, Bryon~A Mueller, Jazmin Camchong, and
  Kelvin~O Lim.
\newblock Altered resting state complexity in schizophrenia.
\newblock {\em NeuroImage}, 2011.

\bibitem{Richiardi2013}
J.~Richiardi, S.~Achard, H.~Bunke, and D.~Van De~Ville.
\newblock Machine learning with brain graphs: Predictive modeling approaches
  for functional imaging in systems neuroscience.
\newblock {\em Signal Processing Magazine, IEEE}, 30(3):58--70, 2013.

\bibitem{leonardi2013principal}
Nora Leonardi, Jonas Richiardi, Markus Gschwind, Samanta Simioni, Jean-Marie
  Annoni, Myriam Schluep, Patrik Vuilleumier, and Dimitri Van De~Ville.
\newblock Principal components of functional connectivity: A new approach to
  study dynamic brain connectivity during rest.
\newblock {\em NeuroImage}, 2013.

\bibitem{schluep2013principal}
Myriam Schluep, Patrik Vuilleumier, and Dimitri Van De~Ville.
\newblock Principal components of functional connectivity: A new approach to
  study dynamic brain connectivity during rest.
\newblock 2013.

\bibitem{fekete2013}
Tomer Fekete, Meytal Wilf, Denis Rubin, Shimon Edelman, Rafael Malach, and
  Lilianne~R. Mujica-Parodi.
\newblock Combining classification with fmri-derived complex network measures
  for potential neurodiagnostics.
\newblock {\em PLoS ONE}, 8(5):e62867, 05 2013.

\bibitem{jietopological}
Biao Jie, Daoqiang Zhang, Chong-Yaw Wee, and Dinggang Shen.
\newblock Topological graph kernel on multiple thresholded functional
  connectivity networks for mild cognitive impairment classification.

\bibitem{jie2012structural}
Biao Jie, Daoqiang Zhang, Chong-Yaw Wee, and Dinggang Shen.
\newblock Structural feature selection for connectivity network-based mci
  diagnosis.
\newblock In {\em Multimodal Brain Image Analysis}, pages 175--184. Springer,
  2012.

\bibitem{feizi2013network}
Soheil Feizi, Daniel Marbach, Muriel M{\'e}dard, and Manolis Kellis.
\newblock Network deconvolution as a general method to distinguish direct
  dependencies in networks.
\newblock {\em Nature biotechnology}, 2013.

\bibitem{zou2005}
Hui Zou and Trevor Hastie.
\newblock Regularization and variable selection via the elastic net.
\newblock {\em Journal of the Royal Statistical Society: Series B (Statistical
  Methodology)}, 67(2):301--320, 2005.

\bibitem{tibshirani1996regression}
Robert Tibshirani.
\newblock Regression shrinkage and selection via the lasso.
\newblock {\em Journal of the Royal Statistical Society. Series B
  (Methodological)}, pages 267--288, 1996.

\bibitem{DeMol2009}
Christine~De Mol, Ernesto~De Vito, and Lorenzo Rosasco.
\newblock Elastic-net regularization in learning theory.
\newblock {\em Journal of Complexity}, 25(2):201 -- 230, 2009.

\bibitem{mosci2010solving}
Sofia Mosci, Lorenzo Rosasco, Matteo Santoro, Alessandro Verri, and Silvia
  Villa.
\newblock Solving structured sparsity regularization with proximal methods.
\newblock In {\em Machine Learning and Knowledge Discovery in Databases}, pages
  418--433. Springer, 2010.

\bibitem{van2009attention}
Marcel van Gerven and Ole Jensen.
\newblock Attention modulations of posterior alpha as a control signal for
  two-dimensional brain--computer interfaces.
\newblock {\em Journal of neuroscience methods}, 179(1):78--84, 2009.

\bibitem{bahramisharif2010covert}
Ali Bahramisharif, Marcel Van~Gerven, Tom Heskes, and Ole Jensen.
\newblock Covert attention allows for continuous control of brain--computer
  interfaces.
\newblock {\em European Journal of Neuroscience}, 31(8):1501--1508, 2010.

\bibitem{signoretto2012classification}
Marco Signoretto, Emanuele Olivetti, Lieven De~Lathauwer, and Johan~AK Suykens.
\newblock Classification of multichannel signals with cumulant-based kernels.
\newblock {\em IEEE Transactions on Signal Processing}, 60(5):2304--2314, 2012.

\bibitem{kia2013discrete}
Seyed~Mostafa Kia, Emanuele Olivetti, and Paolo Avesani.
\newblock Discrete cosine transform for meg signal decoding.
\newblock In {\em Pattern Recognition in Neuroimaging (PRNI), 2013
  International Workshop on}, pages 132--135. IEEE, 2013.

\bibitem{olivetti2010}
Emanuele Olivetti, Andrea Mognon, Susanne Greiner, and Paolo Avesani.
\newblock Brain decoding: biases in error estimation.
\newblock In {\em Brain Decoding: Pattern Recognition Challenges in
  Neuroimaging (WBD), 2010 First Workshop on}, pages 40--43. IEEE, 2010.

\bibitem{breiman2001random}
Leo Breiman.
\newblock Random forests.
\newblock {\em Machine learning}, 45(1):5--32, 2001.

\bibitem{cortes1995support}
Corinna Cortes and Vladimir Vapnik.
\newblock Support-vector networks.
\newblock {\em Machine learning}, 20(3):273--297, 1995.

\bibitem{albanese2012mlpy}
Davide Albanese, Roberto Visintainer, Stefano Merler, Samantha Riccadonna,
  Giuseppe Jurman, and Cesare Furlanello.
\newblock mlpy: machine learning python.
\newblock {\em arXiv preprint arXiv:1202.6548}, 2012.

\bibitem{hamming1950error}
Richard~W Hamming.
\newblock Error detecting and error correcting codes.
\newblock {\em Bell System technical journal}, 29(2):147--160, 1950.

\bibitem{ipsen2002evolutionary}
Mads Ipsen and Alexander~S Mikhailov.
\newblock Evolutionary reconstruction of networks.
\newblock {\em Physical Review E}, 66(4):046109, 2002.

\bibitem{jurman2011biological}
Giuseppe Jurman, Samantha Riccadonna, Roberto Visintainer, and Cesare
  Furlanello.
\newblock Biological network comparison via ipsen-mikhailov distance.
\newblock {\em arXiv preprint arXiv:1109.0220}, 2011.

\bibitem{jurman2012}
Giuseppe Jurman, Roberto Visintainer, Samantha Riccadonna, Michele Filosi, and
  Cesare Furlanello.
\newblock A glocal distance for network comparison.
\newblock {\em arXiv preprint arXiv:1201.2931}, 2012.

\bibitem{cortes2003positive}
Corinna Cortes, Patrick Haffner, and Mehryar Mohri.
\newblock Positive definite rational kernels.
\newblock In {\em Learning Theory and Kernel Machines}, pages 41--56. Springer,
  2003.

\bibitem{vishwanathan2010graph}
SVN Vishwanathan, Nicol~N Schraudolph, Risi Kondor, and Karsten~M Borgwardt.
\newblock Graph kernels.
\newblock {\em The Journal of Machine Learning Research}, 99:1201--1242, 2010.

\bibitem{mokhtari2012decoding}
Fatemeh Mokhtari and Gholam-Ali Hossein-Zadeh.
\newblock Decoding brain states using backward edge elimination and graph
  kernels in fmri connectivity networks.
\newblock {\em Journal of neuroscience methods}, 2012.

\bibitem{vega2013brain}
Sandro Vega-Pons and Paolo Avesani.
\newblock Brain decoding via graph kernels.
\newblock In {\em Pattern Recognition in Neuroimaging (PRNI), 2013
  International Workshop on}, pages 136--139. IEEE, 2013.

\bibitem{zhou2010time}
Shuheng Zhou, John Lafferty, and Larry Wasserman.
\newblock Time varying undirected graphs.
\newblock {\em Machine Learning}, 80(2-3):295--319, 2010.

\bibitem{holme2012temporal}
Petter Holme and Jari Saram{\"a}ki.
\newblock Temporal networks.
\newblock {\em Physics reports}, 519(3):97--125, 2012.

\bibitem{de2013mathematical}
Manlio De~Domenico, Albert Sol{\'e}-Ribalta, Emanuele Cozzo, Mikko Kivel{\"a},
  Yamir Moreno, Mason~A Porter, Sergio G{\'o}mez, and Alex Arenas.
\newblock Mathematical formulation of multi-layer networks.
\newblock {\em arXiv preprint arXiv:1307.4977}, 2013.

\bibitem{sole2013spectral}
Albert Sole-Ribalta, Manlio De~Domenico, Nikos~E Kouvaris, Albert Diaz-Guilera,
  Sergio Gomez, and Alex Arenas.
\newblock Spectral properties of the laplacian of multiplex networks.
\newblock {\em Physical Review E}, 88(3):032807, 2013.

\bibitem{kivela2013multilayer}
Mikko Kivel{\"a}, Alexandre Arenas, Marc Barthelemy, James~P Gleeson, Yamir
  Moreno, and Mason~A Porter.
\newblock Multilayer networks.
\newblock {\em arXiv preprint arXiv:1309.7233}, 2013.

\bibitem{rubinov2010}
Mikail Rubinov and Olaf Sporns.
\newblock Complex network measures of brain connectivity: uses and
  interpretations.
\newblock {\em Neuroimage}, 52(3):1059--1069, 2010.

\bibitem{girvan2002community}
Michelle Girvan and Mark~EJ Newman.
\newblock Community structure in social and biological networks.
\newblock {\em Proceedings of the National Academy of Sciences},
  99(12):7821--7826, 2002.

\bibitem{newman2004finding}
Mark~EJ Newman and Michelle Girvan.
\newblock Finding and evaluating community structure in networks.
\newblock {\em Physical review E}, 69(2):026113, 2004.

\bibitem{porter2009communities}
Mason~A Porter, Jukka-Pekka Onnela, and Peter~J Mucha.
\newblock Communities in networks.
\newblock {\em Notices of the AMS}, 56(9):1082--1097, 2009.

\bibitem{borgatti2000models}
Stephen~P Borgatti and Martin~G Everett.
\newblock Models of core/periphery structures.
\newblock {\em Social networks}, 21(4):375--395, 2000.

\bibitem{holme2005core}
Petter Holme.
\newblock Core-periphery organization of complex networks.
\newblock {\em Physical Review E}, 72(4):046111, 2005.

\bibitem{rombach2012core}
M~Puck Rombach, Mason~A Porter, James~H Fowler, and Peter~J Mucha.
\newblock Core-periphery structure in networks.
\newblock {\em arXiv preprint arXiv:1202.2684}, 2012.

\bibitem{bassett2013task}
Danielle~S Bassett, Nicholas~F Wymbs, M~Puck Rombach, Mason~A Porter, Peter~J
  Mucha, and Scott~T Grafton.
\newblock Task-based core-periphery organization of human brain dynamics.
\newblock {\em PLoS computational biology}, 9(9):e1003171, 2013.

\end{thebibliography}
\end{document}